\documentclass[journal]{IEEEtran}

\usepackage[utf8]{inputenc}
\usepackage{authblk}

\usepackage{amsmath,amssymb,amsfonts}
\usepackage{textcomp}
\usepackage{amsthm}
\usepackage{mathtools}
\usepackage{arydshln} %for the dashed lines in the matrices
\usepackage{dsfont}
\usepackage{lipsum}
\usepackage{wrapfig}
\usepackage{color}
\usepackage{fancyhdr}
\usepackage{cite}
\usepackage{amsmath}
\usepackage{hyperref}
\usepackage{url}
\usepackage{algpseudocode}
\usepackage{algorithm}
\usepackage{amsmath}
\usepackage{bigints}

\usepackage{caption}
\usepackage{tabularx}
\usepackage{subcaption}
\usepackage{graphicx}

\DeclareMathOperator\erf{erf}

\begin{document}
%%%%%%%%%%%%%%%%%
\title{A Novel Approach to Climate Resilience of Infrastructure Networks}

\author{Qianqian Li, 
        \and Giuliano Punzo,
        \and Craig Robson,
        \and Hadi Arbabi, 
        \and Martin Mayfield
        % <-this % stops a space

\thanks{Q. Li, H. Arbabi, and M. Mayfield are with Department of Civil and Structural Engineering, The University of Sheffield}
\thanks{G. Punzo is with Department of Automatic Control and Systems Engineering, The University of Sheffield}
\thanks{C. Robson is with School of Engineering, Newcastle University}

}



\maketitle

\begin{abstract}
With a changing climate, the frequency and intensity of extreme weather events are likely to increase, posing a threat to infrastructure systems' resilience. The response of infrastructure systems to localised failures depends on whether assets are affected randomly, in a targeted strategic way, or any way in between. More than that, infrastructure decisions today, including new routes or improvements to existing assets, will underpin the behaviour of the systems over the next century. It is important to separate and analyse the case of climate-based disruptions and how they affect systems' resilience. This paper presents a probabilistic resilience assessment framework where failure scenarios and network disruptions are generated using weather profile data from climate prediction models with component-level fragility functions. A case study is then carried out to quantify the resilience of Great Britain's railway passenger transport system to high-temperature-related track buckling under the Representative Concentration Pathway 8.5 (RCP8.5) climate change scenario. A 95-year horizon on the resilience of the railway system is drawn. The results also reveal the non-linear responses of the railway system to the increasing temperature and show that models considering random asset failures overestimate the system's resilience. 
\end{abstract}

%%%%%%%%%%%%%%%%%%%%%%%%%%%%%%%%%%%%%%%%%%%%%%%%%%%%%%%%%%%%%%%%%%%%%%%%%%%%%%%%%%%%	
\section{Introduction} \label{sec:1}
\subsection{Background}
The centrality of infrastructure systems in society, and therefore their resilience to disruptions in a complex, fast-evolving environment, are now universally recognised \cite{Rinaldi2001,Chester2021,Saidi2018}.
Extreme weather events such as storms and floods can extensively affect the functionality and serviceability of infrastructure systems \cite{panteli2015influence,ferranti2016heat,pant2018critical,balomenos2018fragility}. The current climate change trajectory is only likely to result in an increased frequency and intensity of such events \cite{IPCC2013,AR6Chp11}. Although current infrastructure systems have been stressed by various types of events from time to time, they are generally considered to be resilient to specific natural hazards as they are designed, built and operated in compliance with design codes and regulations set on historical meteorological data \cite{eu-circle_eu-circle_2018}. Yet, design standards and operational standards, and therefore the system's capacity to absorb shocks, are defined over expected magnitudes of shocks. Such expectations have been largely surpassed by the scale of extreme weather events caused by climate change \cite{met_office_2019, finlay_autumn_2022}. Therefore, current infrastructure systems may not have the ability to withstand the future climate, characterised by more frequent and intense weather extremes. What is resilient to the present-day climate may be vulnerable to the future climate. An understanding of how weather hazard impacts infrastructure systems is required.

Since the publication of the Intergovernmental Panel on Climate Change (IPPC) Fifth Assessment Report \cite{IPCC2013} and the open access to climate model output data, there has been an increasing body of literature on assessing the impacts of climate change on infrastructure systems. For example, the Climate Change Risk Assessment 2017 \cite{sayers2015climate} assesses the risk of future flood risk to Great Britain and generates flood risk maps under three climate change scenarios. The number of assets exposed is estimated by overlapping the flood risk maps with the geographical maps of the assets and identifying those likely to surpass specific indices or risk thresholds. Other impact assessments appearing in the literature have been carried out in a similar manner. They rely on estimating the likelihood of surpassing certain design or operational thresholds under several climate change scenarios \cite{matulla2018climate, sanchis2020risk, sutherland2002coastal, bubeck2019global, mccoll2012assessing}. 

For the classification and determination of the impact of climate hazards, a two-tier approach is proposed in \cite{eucircle2018D3.5}, where the approach first distinguishes direct and indirect impact. Direct impacts refer to consequences related to the infrastructure system itself, including complete or partial damage to physical infrastructure assets, deviation of performance from the fully functional level, and connectivity loss. Indirect impacts refer to those received by the society that is served by the infrastructure systems, such as causalities, community isolation (both physical and in terms of communications), and economic losses. By this classification, the majority of existing climate change impact assessments, including \cite{sayers2015climate, matulla2018climate, sanchis2020risk, sutherland2002coastal, bubeck2019global, mccoll2012assessing}, are on the physical damage level. 

Under the scope of climate change impact assessments, functional damages are rarely analysed, often just viewed as a consequence. Works related to functional damage and system-level loss of service, in particular, are comparatively more limited in number \cite{mostafavi2018system, bollinger2016evaluating}. Limiting impact assessments to the estimation of physical damages, or even component-level functional loss overlooks the complex interdependencies of infrastructure systems \cite{Rinaldi2001}. Infrastructure systems are well acknowledged as complex coupled systems, the behaviour or response of which is distinct from the combined behaviour or response of its components \cite{punzo_challenges_2018}. Impact assessments on physical damage level certainly provide valuable insights into the magnitude of climate-change-related disruptions. However, the complex dynamics characterising infrastructure systems' responses are such that system-level effects cannot be derived in a straightforward manner from the component level.
Therefore, an understanding of how more extreme weather events, including those not normally seen in geographic areas in the past, may cause a hazard and impact infrastructure systems is required. 

\subsection{State of the Art}
Existing works on system-level response to disruptive events are often cast within research in resilience, as well as the related areas of vulnerability and robustness. A resilience assessment often starts with modelling real-world infrastructure systems as networks. Network models in literature can be categorised into two groups: topological models and flow models. In a topological model, e.g., \cite{von2009public}, physical infrastructure system assets, such as rail stations, transmission lines, and airports, are modelled as nodes or edges in the network. These models emphasise the topological structure of the system but lack the ability to capture the functional aspect of the infrastructure system: where the demands are, where the supplies are and how any demand can be met. A flow model, e.g., \cite{holden2013network}, has its emphasis on the services and flows delivered by the infrastructure system more than the network’s topological structure. Flow models are normally expressed as OD (origin-destination) matrices. In a flow model, nodes are entities that either supply, demand, or transit services or goods. Edges capture flows between pairs of nodes. Recent works on system-of-systems, interdependent, or interconnected networks adopt a combined topological and flow model (e.g. \cite{thacker2017system, pant_vulnerability_2016, goldbeck2019resilience}). Such models have separated asset and flow layers representing the physical infrastructure assets as a graph and the services provided, respectively. Inter-layer dependencies describe the physical embedding of a service end-node into the asset layer, that is, in which asset node a flow between two nodes of the flow layer is originated or delivered by means of the physical network of assets. 

Disruptions are mostly simulated as strategic removals of network components \cite{zhang2015assessing, berche2009resilience, pagani2019resilience, holmgren2006using}. This type of network component removal is often referred to as an \textit{attack} in abstract network studies \cite{Albert2000}. The most common strategies used for network attacks are either random, where nodes and edges are randomly selected and removed from the network, or targeted, where nodes and edges are selected based on their structural/topological importance in the network \cite{gallos2005stability}. The random attack strategy resembles some real-world disruption events like random equipment failure, operational faults, and accidents \cite{pant_vulnerability_2016}. The targeted attack strategy, to some extent, aims at capturing events like terrorist attacks \cite{huang2011robustness} or some theoretic worst case scenarios. While random disruptions and targeted attacks are useful simplifications, they do not completely cover the wide variety of possibilities that a real-world scenario may present. Many disruptive events, particularly weather-related events, may not fit into either. 

Although weather events feature stochasticity, they are not purely random because the climate has deterministic dynamics that exhibit chaotic behaviours \cite{elsner1992nonlinear}. However, weather-related disruptions certainly do not maliciously target any specific network components, making the targeted attack strategy unrealistically severe. Those approaches are not capable of capturing the feature of weather-related disruptions. Furthermore, complex networks behave differently under different attack scenarios \cite{berche2009resilience}. Even if the infrastructure network of interest is shown to be resilient to random or targeted network attack strategies, it is not necessarily resilient to climate, or extreme weather event-based disruptions. Therefore, the resilience of infrastructure systems to weather-related events should be simulated with realistic weather profiles in addition to the random and targeted strategies. 

Several studies have attempted to use weather profiles to initiate network disruptions. Panteli and Mancarella \cite{panteli2015influence} propose a conceptual framework to assess the influence of climate change on weather-related power interruption. In their work, an explicit reference is made to the use of weather profile data in both time and space domains to initiate system component failure and simulate cascading effects. However, subsequent works appear not to implement such a strategy fully. The hourly wind profiles can be obtained by sampling these probability distributions. When assessing power network resilience, which is done using scaled-up time-series wind profiles generated from a weather simulator \cite{panteli_power_2017}. Works that account for actual spatial weather patterns, e.g., \cite{fu2018integrated}, do so by reproducing historical wind extremes with spatial correlation. This implies the assumption that historical or present-day weather statistics will hold in the future, which is now widely recognised as a fallacy due to the changing climate and weather patterns \cite{IPCC2013, AR6Chp11}. What is resilient to the current patterns may easily be fragile to the future ones. Future weather profiles produced by climate models are the only viable option to assess the resilience of infrastructure systems to future extreme weather events, moving beyond the widespread extremes of random and targeted attacks, unable to capture the features and threats of climate change. 

\subsection{Main Contributions of the Work} 
This paper proposes a systemic approach for assessing the resilience of infrastructure systems to climate change. In doing so, it proposes a novel quantitative framework that returns the statistical distribution of the system's response as produced by weather profile data for the relevant geographical area. These are obtained from climate model outputs and projected onto the geographical asset location of an infrastructure system. Moving beyond the current approaches to quantify infrastructure network resilience, this work

\begin{itemize}
    \item proposes a method to initiate network disruption based on local climate hazards obtained from weather profile/data with a 95-year horizon;
    \item assesses system-level functional loss based on the service level retained, which considers possible reconfiguration, as opposed to a mere count of the failed nodes or the identification of a threshold for network fragmentation.
    \item brings the physical damage and service loss quantification together to return a measurement of the infrastructure systems resilience to extreme weather events-based failure scenarios.
    \item benchmarks the method through a case study on Great Britain's railway network, including disruptions to the ability to satisfy the travel demand.
\end{itemize}
Proceeding sections are organised as follows. Section \ref{sec:2} describes key components in the resilience assessment framework, offering a clear account of the formulation proposed, which can be replicated for further research or adapted to specific infrastructure systems. In Section \ref{sec:3}, the proposed framework is applied to Great Britain's railway network with high-temperature-related failures. Discussion and conclusions are offered in Section \ref{sec:4} and Section \ref{sec:6}.

%%%%%%%%%%%%%%%%%%%%%%%%%%%%%%%%%%%%%%%%%%%%%%%%%%%%%%%%%%%%%%%%%%%%%%%%%%%%%%%%%%%%
\section{The Resilience Assessment Framework}\label{sec:2}
\begin{figure}[h] 
\begin{center} 
	\includegraphics[width=1\columnwidth]{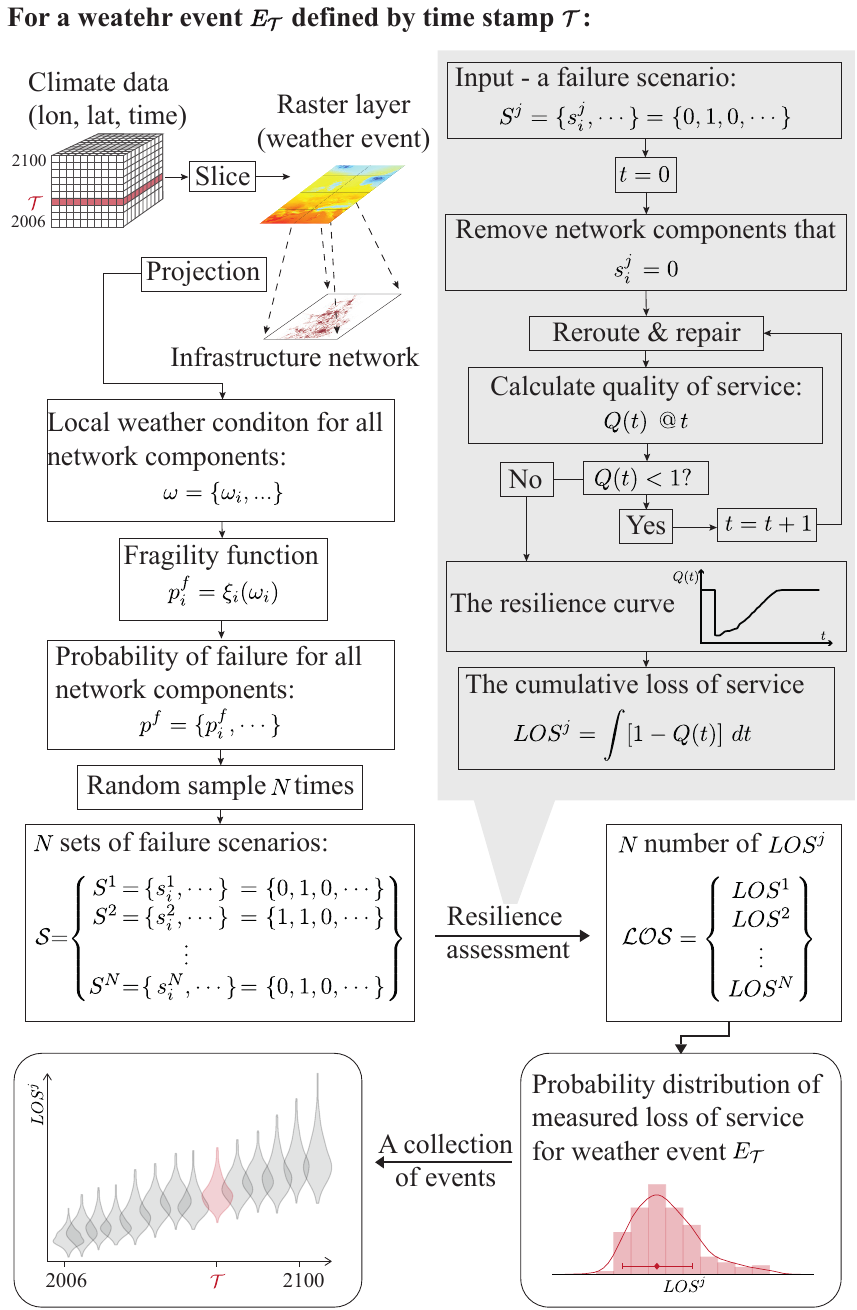}
\end{center}
\caption{Overview of the probabilistic resilience assessment framework.}
\label{fig: framework}
\end{figure}

The proposed method uses standard weather profiles obtained from the Earth System Grid Federation, which holds the most extensive collection of observational, reanalysis, and simulation data for climate change research. Those weather profiles are available as individual time series of a weather parameter, such as wind, precipitation, and humidity, with different time resolutions and geographical ranges. The key elements of the proposed approach, namely the network model, the failure scenario generation, and the probabilistic resilience measure, are described in more detail later in this section. It is worth mentioning that resilience in this work does consider the recovery stage, where rerouting and repairing activities are concerned. However, the recovery algorithm is not presented in this section as they could be interchangeable depending on the case study or the infrastructure sector being analysed. The high-level architecture of the proposed framework is shown in Figure \ref{fig: framework}.

\subsection{Network Representation of Infrastructure Systems}\label{sub:Network_rep}
In this work, the infrastructure system is modelled as a bi-layer network, constituting an asset and a flow layer. The reason for this is twofold. Provided the ultimate purpose of an infrastructure system is to provide a service or services, a resilience measurement that relates to the system’s ability to maintain the delivery of services when exposed to external shocks would be better than one relating to the extent of damage/change on its topological structure. The second reason is that such separation makes it possible to incorporate multiple asset layers and therefore multiple types of climate hazards into the assessment. The successful delivery of a unit of service can sometimes rely on multiple interdependent infrastructure systems with components in each system affected by different types of climatic hazards. The separation makes it possible to initiate failures in different asset layers caused by different climate hazards simultaneously while increasing the flexibility of the modelling framework to adapt to different infrastructure systems. 

In the following, the asset and flow layers will be indicated by subscripts $\alpha$ and $\phi$, respectively. Hence, for the asset layer, assets can be thought as a graph $\mathcal{G}_{\alpha}=\{\mathcal{V}_\alpha, \mathcal{E}_{\alpha}\}$ where $\mathcal{V}_\alpha$ and $\mathcal{E}_\alpha$ indicate the set of nodes and edges, respectively. The framework does not need to differentiate node assets and edge assets rigidly. For layer $\alpha$, $\mathcal{A}$ indicates the set of assets in layer $\alpha$, with elements $\mathcal{A}_i,~ i = 1,2,\cdots,.|\mathcal{V}_{\alpha}|+|\mathcal{E}_{\alpha}|$, with $|\cdot |$ indicating the cardinality of a set. The equivalent definitions for the flow layer $\phi$ are omitted for brevity.

\subsection{Failure Scenario Generation}\label{sub:attacks}
In this work, a \textit{failure scenario} refers to a set consisting of the simultaneous failure and removal of network components. The generation of such failure scenarios follows a systematic approach using the obtained climate model output. The climate outputs are three-dimensional (3D) data, where spatial weather data are combined with the third dimension of time. A \textit{weather event} refers to a slice from the 3D data with a desired geographical range and timescale. For a given weather event, by projecting the weather data on the asset layer, the local weather condition for all network components in the asset layer can be found:
\[\mathbf{\omega} = \{\omega_i, ...\},~ i = 1,2,...|\mathcal{V}_{\alpha}|+|\mathcal{E}_{\alpha}|\]
where $\omega_i$ is the local weather condition for asset $\mathcal{A}_i$. 

\begin{figure}[h]
\begin{center} 
	\includegraphics[width=0.95\columnwidth]{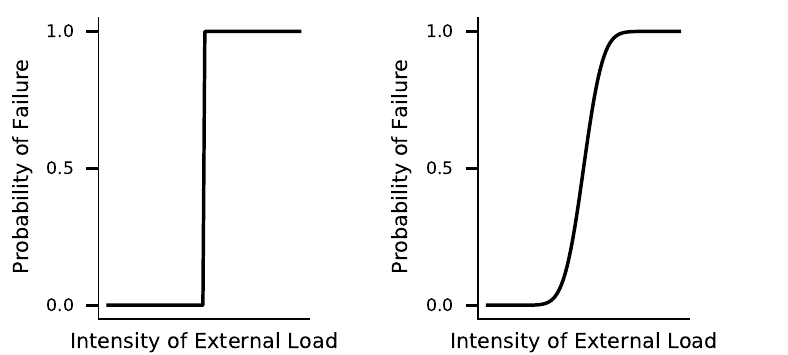}\\
	(a)\hspace{4cm} (b)
\end{center} 
\caption{Example fragility curves: a) step function form; b) sigmoid function form} 
\label{fig: fragilility}
\end{figure}

With the local weather condition allocated for each asset network component, a non-zero asset failure probability arises as the asset may fail due to local weather conditions. This probability defines the likelihood that an asset can withstand an assigned local weather condition and can be derived from a fragility function, which returns the probability of failure as a function of the magnitude of the local weather conditions. The shape of a fragility curve reflects uncertainty in the asset’s ability to withstand a shock. If the failure is deterministic (e.g., a  circuit breaker in an electric circuit triggered by a given value of the current), the fragility function takes the shape of a step function with a threshold (Figure \ref{fig: fragilility}a), beyond which the probability of failure passes abruptly from 0 to 1. Examples used in the literature are air temperature threshold for railway track buckling \cite{sanchis2020risk}, depth of flooded water on railway tracks \cite{sutherland2002coastal}, and significant wave height on port operation \cite{camus2019probabilistic}. The threshold model, or step function, assumes that asset failures are deterministic when the assigned shock is above the threshold. When there are greater uncertainties in the asset’s capacity to withstand a shock, a more general sigmoid function can be used. Examples of the latter include \cite{panteli_modeling_2017, panteli_power_2017, fu2018integrated}, where sigmoidal fragility functions are associated with the failure of electricity transmission lines and towers to local wind speed.

In the most general form, consider the fragility function for each network components in the asset layer as
\begin{equation}
    p_i^f = \xi_i(\omega_i)
\end{equation}
where $\xi_i$ stands for the fragility function for asset $\mathcal{A}_i$ and $p_i^f$ is the calculated probability of failure for asset $\mathcal{A}_i$ when the local weather condition is $\omega_i$. The probability of failure for all network components in the asset layer is then
\[p^f = \{p_i^f, \cdots\}, ~ i = 1,2,...|\mathcal{V}_{\alpha}|+|\mathcal{E}_{\alpha}|\]\quad
For the sake of simplicity, and without loss of generality, all assets are assumed to have binary state. The state of a single network component, $\mathcal{A}_i$, can therefore be described by an independent Bernoulli random variable $X_i \sim Ber(p_i^f)$, so that
\begin{equation}
  P(X_i)=\begin{cases}
    p_i^f, & \text{for $X_i = 0 $}\\
    1-p_i^f, & \text{for $X_i = 1 $}.
  \end{cases}
\end{equation}
The state of the whole network is therefore effectively controlled by a set of independent Bernoulli random variables
\[\mathcal{X} = \{X_i, \cdots\}, ~ i = 1,2,...|\mathcal{V}_{\alpha}|+|\mathcal{E}_{\alpha}|\]
each corresponding to some failure probability. 

One sample of each random variable returns a failure scenario for the system, where network component failures are initiated by removing assets with $X_i = 0$. Each failure scenario represents one possible outcome from the weather event. The complete sample space for $\mathcal{X}$ is of size \(2^{|\mathcal{V}_{\alpha}|+|\mathcal{E}_{\alpha}|}\). %This includes the worst possible scenario where \(X_i=0\)  for all network components and the best scenario where \(X_i=1\) for all network components.
To reduce the complexity of simulating the system’s response to all \(2^{|\mathcal{V}_{\alpha}|+|\mathcal{E}_{\alpha}|}\) combinations, a Monte Carlo approach is used to estimate the possible outcomes. Through repeated sampling, $N$ set of possible failure scenarios can be obtained: 
\[\mathcal{S} = \{S^j, \cdots \}, ~ j = 0, \cdots, N\]
where $j$ denotes a single Monte Carlo run. Each subset, $S^j$, contains a combination of states for all network components and is regarded as a single failure scenario. 
\[S^j = \{s_i^j, \cdots \}, ~ i = 1,2,...|\mathcal{V}_{\alpha}|+|\mathcal{E}_{\alpha}|\]

\subsection{Probabilistic resilience measure} \label{sub:resilience_mes}
This paper considers resilience as the ability of a system to maintain or return to its normal operations after a disruption occurs. This derives from an ecological perspective and is first introduced by Holling \cite{holling1973resilience}. However, various definitions and measures of resilience have been proposed in applications to systems of different natures \cite{hosseini2016review}. Here the resilience metric proposed by Bruneau et al. \cite{bruneau2003framework} is adopted for its general applicability. It measures resilience as cumulative service degradation from the time of earthquake, or any disruptive events, happening, \(t_0\), to the time of full recovery, \(t_1\), as shown in the equation \ref{eq_R}. \(Q(t)\) denotes the quality of infrastructure at time \(t\). 
\begin{equation} \label{eq_R}
    R = \int_{t_0}^{t_1} [100-Q(t)]~ dt
\end{equation}
In this work, the resilience metric from \cite{bruneau2003framework} is modified in two aspects. First, instead of the quality of the infrastructure assets, \(Q(t)\) here measures the quality of service provided. Moreover, such a measure of the quality of services may take different meanings depending on the nature of the infrastructure system under study. As a general case, it is expressed as the percentage of satisfied demand:
\begin{equation} \label{eq_Qt}
    Q(t) = \frac{\sum_{\phi} F^t}{\sum_{\phi} D^t}
\end{equation}
where $F^t$ and $D^t$ denote the delivery and demand at time $t$ respectively. 
The resilience metric is then calculated as cumulative loss of service, \textit{LOS}: 
\begin{equation} \label{eq_los}
    LOS = \bigintsss_{t_0}^{t_1}{ 1-\left(\frac{\sum_{\phi} F^t}{\sum_{\phi} D^t} \right) }dt
\end{equation}

Second, to take the uncertainty of the system behaviour into account, resilience to a given weather event is associated with the averaged LOS from all sampled failure scenarios. For a given weather event, a set of $N$ failure scenarios, $\mathcal{S} = \{S^j, \cdots \}, ~ j = 0, \cdots, N$, are randomly sampled. For each failure scenario, \(S^j\), component failures are initiated in the asset layer accordingly and disruptions in the service layer are then computed. Depending on the system's dynamics, reconfiguration and repair activities can then be performed. \(Q(t)\), is calculated at every time step until full service is recovered. The cumulative loss of service, \(LOS_j\), from the $j$-th sampled failure scenario, \(S^j\), is then calculated with Equation \ref{eq_los}. The statistical distribution of set, $\mathcal{LOS} = \{LOS_j\}, ~ j = 0, \cdots, N$, describes the system's resilience to the given weather event.

%%%%%%%%%%%%%%%%%%%%%%%%%%%%%%%%%%%%%%%%%%%%%%%%%%%%%%%%%%%%%%%%%%%%%%%%%%%%%%%%%%%%
\section{Case Study - Great Britain's Railway system}\label{sec:3}
This section presents a case study on Great Britain's railway passenger transport system, which is modeled as a flow network dependent on a single asset layer of train tracks subjected to high-temperature related track buckling. A set of temperature projections covering the years 2006-2100 from the European Domain of Coordinated Downscaling Experiment (EURO-CORDEX) on a 0.11\textdegree spatial resolution under the Representative Concentration Pathway 8.5 (RCP8.5) climate change scenario is used to generate plausible climate-based failure scenarios. Those failure scenarios are then used to initiate failures in the network, followed by assessments of the resilience to these scenarios.

\subsection{Model Inputs and Assumptions} \label{sec:3a}
In this case study, a few simplifications and assumptions are made for the balance of generality and specificity. In the absence of reliable information and empirical data, those choices are the most general and possible form. 

\subsubsection{The network}
\begin{figure}[h]
\begin{center} 
	\includegraphics[width=0.95\columnwidth]{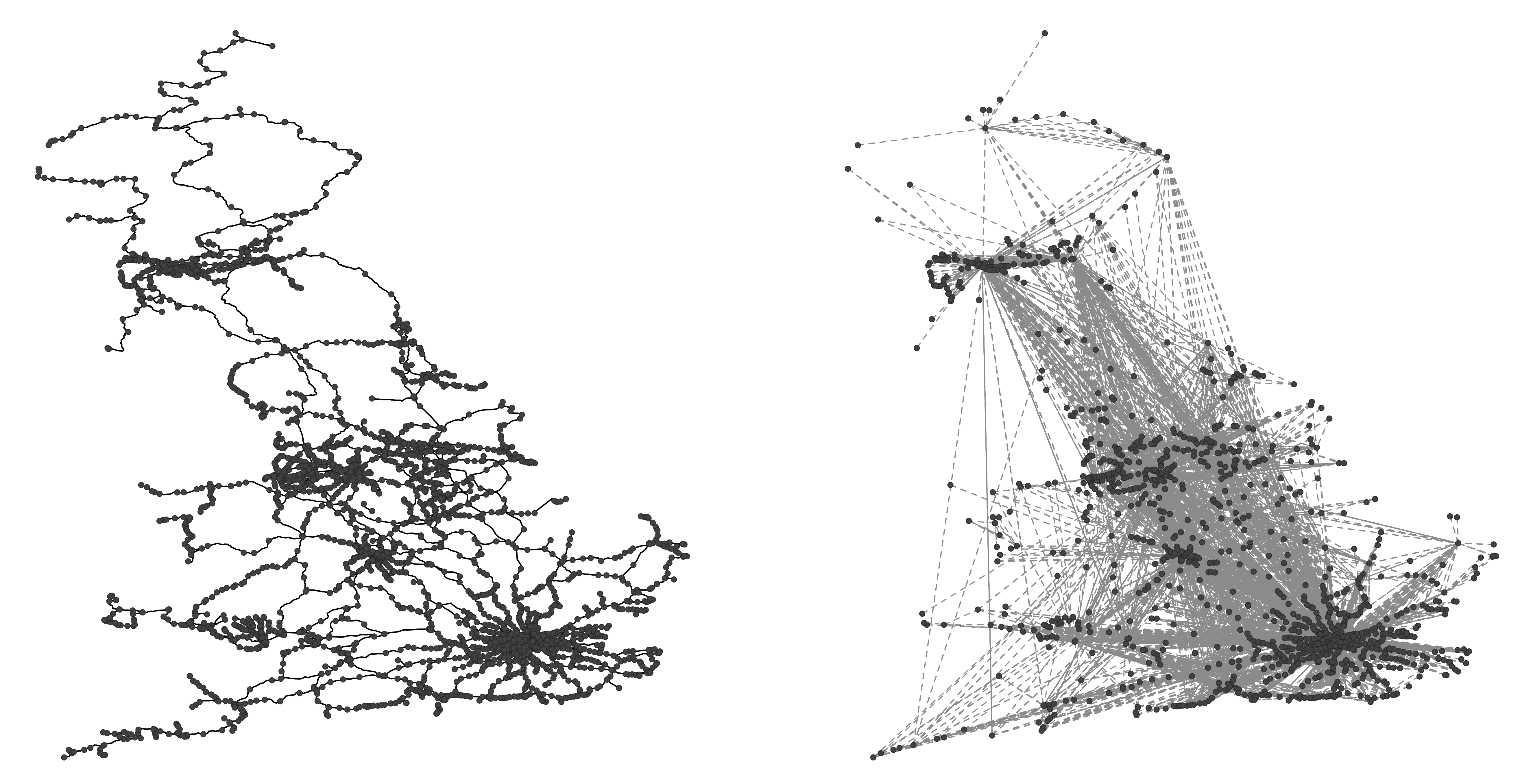}\\
	(a)\hspace{4cm} (b)
\end{center} 
\caption{(a) Asset layer of the network model with 4024 nodes and 4524 edges. (b) Flow layer of the network model. This figure only plots 22,326 out of the total 2,282,270 OD pairs, which has more than 15 passenger trips and geographical path length greater than 30km.} 
\label{fig: network_map}
\end{figure}

The railway network model developed by Pant et al. \cite{pant_vulnerability_2016} is used in this case study. It has separated flow and asset layers (Figure \ref{fig: network_map}). The asset layer is an undirected weighted network that consists of 4024 stations, modeled as nodes; and 4524 railway track segments, modeled as edges. The flow layer is in the form of OD trips, representing the services provided. There are 2,282,270 OD pairs between 2484 origin and destination nodes out of the 4524 nodes in the asst layer. Each OD pair has an original edge path assigned, $\mathbb{P}_{od}^0$, detailing which edges in the asset layer are utilized. Average daily traffic over the year is used as both a measure of the traffic volume on edges in the asset layer and OD demand. The system is assumed to be in a steady state with this daily traffic with the same amount of services demanded every day, regardless of the state of the asset layer. Any daily, weekly, or seasonal variations are not considered in this case study. A time resolution of one day is used.

\subsubsection{The Hazard}
High-temperature-related track buckling is used as an example hazard to demonstrate how to generate network failure scenarios using weather profiles. From \cite{palin2013future, ferranti2016heat}, track buckling is related to the daily maximum temperature and the critical air temperature for heat-related track buckling varies depending on track condition. However, track conditions are not available from the network model or network-wide from other sources. For the purpose of this case study, the dependence of the critical air temperature on track condition \cite{palin2013future} is therefore not considered and all edges in the network are subjected to the same fragility function regardless of any variations in condition. To the best of found knowledge, there is no existing fragility function that expresses the probability of track buckling as a function of the ambient temperature. The empirical definition of such a function is beyond the scope of this paper. A Gaussian sigmoid is hence considered as a fragility function with its shape controlled by the two parameters \(\mu\) and \(\sigma\).  Taking the thresholds in \cite{palin2013future, vajda2014severe} into consideration, the cumulative distribution function of \(\mathcal{N}(35, 2.5)\) is used. The value of \(\mu\) and \(\sigma\) can be tuned if more information becomes available, e.g., a collection of historical failure events and corresponding local temperature. Therefore, 
\begin{equation}\label{eq_frag_CDF}
    p_i^f = \xi(\omega_i) = \frac{1}{2} [1 + \erf(\frac{\omega_i-\mu}{\sigma \sqrt{2}})]
\end{equation}

%Palin (2013) calculates the critical air temperatures (CAT) for the implementation of different speed restriction in relation to track buckling using Network Rail’s specification documents and equations from Hunt (1994). For example, the CAT for an undisturbed, fully ballasted and consolidated track for watchman-on-site, 30/60mph speed restriction, 20mph speed restriction is 35 \textdegree, 39 \textdegree, 42 \textdegree respectively. Speed restrictions are actions taken over concern of track buckling. To some extent, the lower the speed restriction, the higher the probability of track buckling. Ferranti uses industrial fault data to analysis heat related failures on Southeast England’s railway system. Their study shows the median of ambient maximum air temperature at which track failed is 25 \textdegree with an interquartile range of 5.8 \textdegree. Vajda 2014 suggests 25 \textdegree, 32 \textdegree and 43 \textdegree as the three thresholds for three impact level. 
%I used 35 degree C as an average of 32 and 39. (32 from (25-32-43) in Vajda 2014, 39 from (35-39-42) in Palin 2013) . I excluded ferranti, as the 25+/- 5.8 is conditioned on asset failed. Also, there probably are not many days with temperature higher than 30 in their data collection period. 

\subsubsection{The Climate Data}
As mentioned above, track buckling is related to the daily maximum temperature. Therefore the output variable, daily maximum near-surface air temperature, (\textit{tasmax}) ), is used in this case study. EURO-CORDEX provides climate change data for the European domain, covering a period of 95 years, from 2006 to 2100. By limiting the results to
\begin{itemize}
    \item “domain” = “EUR-11” (0.11\textdegree, \(\sim\)12km spatial resolution)
    \item “experiment” =  “RCP8.5”
    \item “time frequency” = “day”
    \item “variables” = “tasmax” 
\end{itemize}
64 sets of model output are left. They differ in the global climate model used for downscaling, the climate model ensemble, and the regional climate model used.\footnote{The most recently updated one, EUR-11\_CNRM-CERFACS-CNRM-CM5\_rcp85\_r1i1p1\_MOHC-HadREM3-GA7-05\_v2, is used in this case study.}  \\

\subsubsection{Rerouting Algorithm}
Railways usually are not delivering at their maximum capacity \cite{ ramdas_options_2017 }. When the network is partially damaged with some OD pairs losing their original path, the spare capacity allows rerouting to utilise the remaining assets to attempt to deliver those interrupted OD flows. When the original path, \(\mathbb{P}_{od}^0 = \{e_{ij}, \cdots\}\) is interrupted due to one or more edge failures, passengers are assumed to stay at node \(v_o\) and no passenger is waiting at any intermediate nodes on the original path. Any edge, if undamaged, is assumed to have a spare capacity of 50\% of its regular traffic for the rerouting. 
A modified minimum-cost maximum-flow algorithm based on the Edmonds-Karp algorithm \cite{edmonds1972theoretical} is used for the flow assignment here. Considering the size of the OD matrix and the time taken to search for a path in a network of great size, the algorithm is only set to search for the first five\footnote{This is an arbitrary value set on the absence of relevant information.} shortest paths instead of until no path exists. Further, the 'cost' here is the geographical length of the path instead of the number of steps (edges traversed). The algorithm searches for paths in the order of the shortest path that, 1) has available capacity; and 2) is within twice\footnotemark[\value{footnote}] the geographical length of the original path, until no such path can be found. More details about the rerouting strategy are available in the Appendix.

\subsubsection{Recovery}
Asset repairing activities are assumed to take place at every time step until the asset layer is fully recovered. For the sake of this example and to bypass the scarcity of information on the recovery of incidents in the railway sector, damaged edges are subject to the same recovery probability of 0.5 at every time step until fully recovered.

\subsection{Simulation} \label{sec:3b}
As the temporal resolution for the tasmax is daily, a single day is regarded as a weather event, which is assumed uncorrelated to conditions of the previous and following days. This case study does not consider the effects of continuous hot days or heatwaves. For each weather event, the local conditions for each edge are first assigned and transferred to the probability of failure using the fragility function. Then 250 sets of failure scenarios are generated through random sampling. 

For the $j$-th failure scenario, disruptions are initiated through the removal of edges for which $s_i^j = 0$ $i = 1,2,...|\mathcal{V}_{\alpha}|+|\mathcal{E}_{\alpha}|$ at time step zero. Interrupted OD pairs, whose original path $\mathbb{P}_{od}^0$ includes any removed edges, are then identified. The rerouting algorithm then calculates the amount of OD flows rerouted. The total amount of delivery at this time step is the sum of the rerouted and the uninterrupted OD trips. The quality of service, $Q(t)$, is calculated with Equation \ref{eq_Qt}. Undelivered OD flows, partially or completely, at the current time step, plus the steady-state daily demand (the annual average daily traffic) becomes the demand for the next time step. Asset repair is carried out between time steps until all edges are recovered. If the original path of an OD pair is recovered at a time step, from the next time step, any accumulated undelivered trips will be delivered via its original path with the 50\% spare capacity fully utilised instead of rerouting.   

\begin{figure}[h]
        \begin{center} 
        	\includegraphics[width=1\columnwidth]{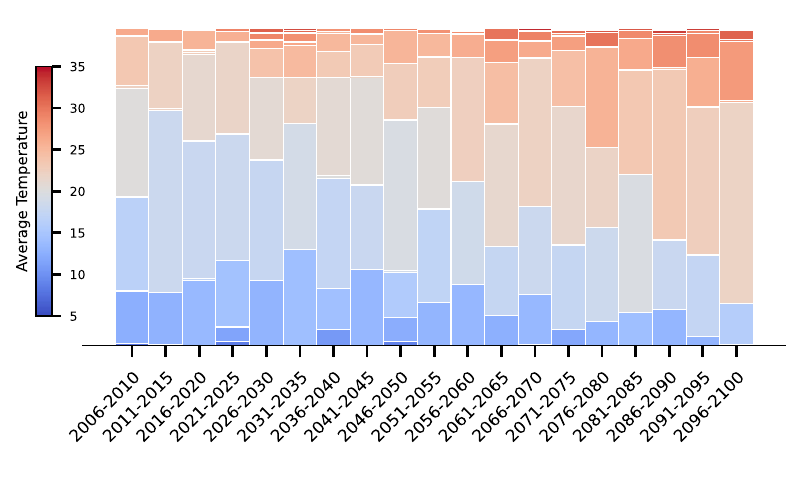}
        \end{center} 
        \caption{Clustering of the summer days (01 May to 30 September) from 2006 to 2100 based on the tasmax of CORDEX output data EUR-11\_CNRM-CERFACS-CNRM-CM5\_rcp85\_r1i1p1\_MOHC-HadREM3-GA7-05\_v2. 10 clusters are produced for each 5-year group through a K-means algorithm. The height of each bar is controlled by the number of days that fall into the cluster. The colour of the bar is controlled by the average temperature of the synthetic centroid of the cluster.}
        \label{clustering}
    \end{figure}

A day-by-day resilience assessment for the 95-year would hardly meet practical considerations and require substantial computational power. A clustering analysis is therefore carried out using a K-means method to categorise each five year’s of summer days, which are most likely to exceed temperature thresholds, into 10 clusters (Figure \ref{clustering}). Euclidean distance is considered and the day closest to the synthetic centroid of each cluster is chosen as the representative day. By doing so, 190 example days are selected, representing 190 typical weather events and thus 190 unique distributions of climate hazards over the railway network. This approach makes the dataset computationally tractable with the method presented without losing significant information.  

To compare the climate-based failure scenarios to random and targeted, shocks of the same intensity should be applied to the network in two separate settings with either a random or targeted strategy. This 'intensity' is taken as the number of edges removed in this work. For a given day, the expected number of failed edges can be calculated as:
\begin{equation}\label{eq:Psi}
    \Psi=E(\mathcal{X}) = \sum_{i=1}^{|\mathcal{E}_{\alpha}|} E(X_i) = \sum_{i=1}^{|\mathcal{E}_{\alpha}|} p_i^f \quad .
\end{equation}
This implies that, if a weather event was set as a random attack on the network assets, on average, $\Psi$ edges would be randomly selected and removed in each Monte Carlo run. For the targeted strategy, edges with the most traffic are removed first. In both cases, the same resilience assessment procedure as in the climate-based failure scenarios is followed.  

%%%%%%%%%%%%%%%%%%%%%%%%%%%%%%%%%%%%%%%%%%%%%%%%%%%%%%%%%%%%%%%%%%%%%%%%%%%%%%%%%%%%%%%%%%%%%%%%%	

\subsection{The 95-year trend}
\begin{figure}[h]
    \centering
    \includegraphics[width=0.95\columnwidth]{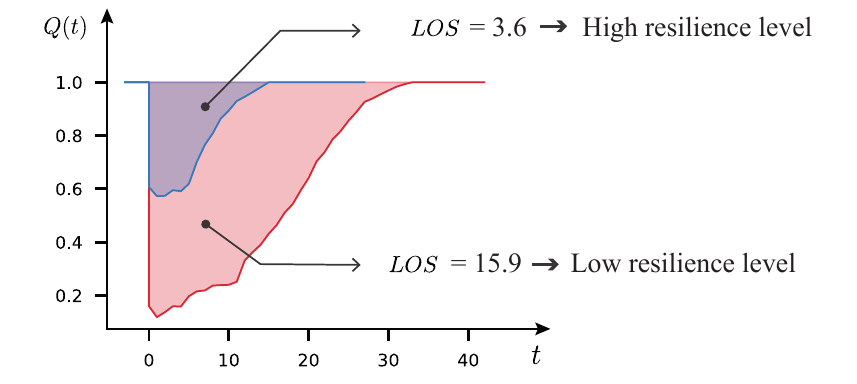}
    \caption{Resilience curve with calculated LOS}
    \label{fig:curve_fill}
\end{figure}

For any given failure scenario, with the resilience defined as in Section \ref{sec:3}, the quality of service drops immediately upon removal of edges. With rerouting and repairing efforts, the quality of service gradually bounces back to one. This forms a system response curve. The area between the curve and the normal performance line ($Q(t)=1$)  forms a triangle, the area of which, effectively is the cumulative loss of service and indicates how resilient the system is (see Figure \ref{fig:curve_fill}. For a given day, 250 failure scenarios are sampled. Therefore, each failure scenario corresponds to a resilience curve and a LOS. The distribution of the 250 LOS shows the likelihood of the outcomes (loss of service) and indicates the system's resilience. A high LOS value means a high degree of service loss overall and is therefore associated with low resilience.

\begin{figure}[h]
    \centering
    \includegraphics{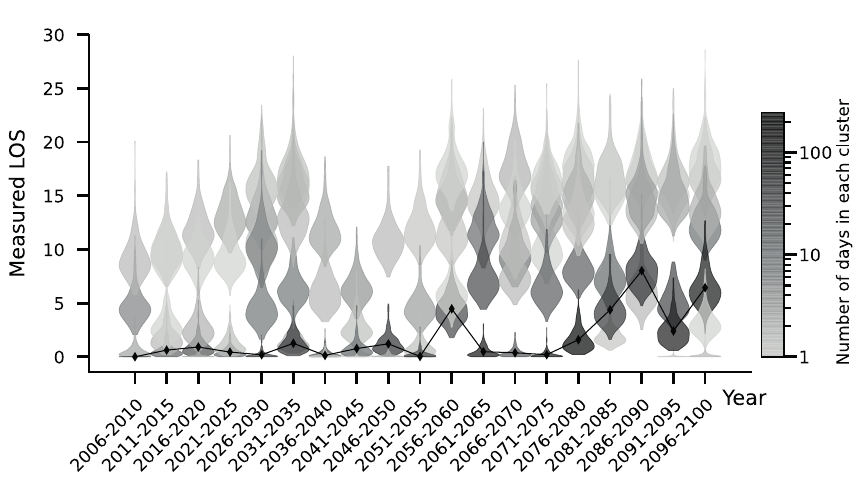}
    \caption{Distribution of the measured cumulative service loss, LOS, of the centroid day of each cluster. Each violin plot shows the distribution of 250 measured LOS from failure scenarios sampled from a single day. This figure only plots days whose expectation of the number of failed edges is greater than 1. The line plot with diamond markers shows the mean of the distribution of the largest cluster.}
    \label{fig:violin_plots}
\end{figure}

To visualise the trend over the 2006-2100 period, these probability density distributions are plotted vertically as violin plots (Figure \ref{fig:violin_plots}). As mentioned above, clustering analyses are carried out to select 10 example days from the summer days of every 5-year period and 190 example days are selected in total. Figure \ref{fig:violin_plots} shows the simulation results of 105 days out of the 190 days as it only includes days whose expected number of failure, $\Psi$, is greater than 1. The colours of the violin plots indicate the number of days in the cluster that the example day belongs to. The darker the plot, the more frequent this day is in the 5-year period it belongs to. From Figure \ref{fig:violin_plots}, the dominant clusters (clusters with the biggest number of days), tends to be associated with a higher degree of LOS towards the end of the century. 

\begin{figure}[h]
    \centering
    \includegraphics{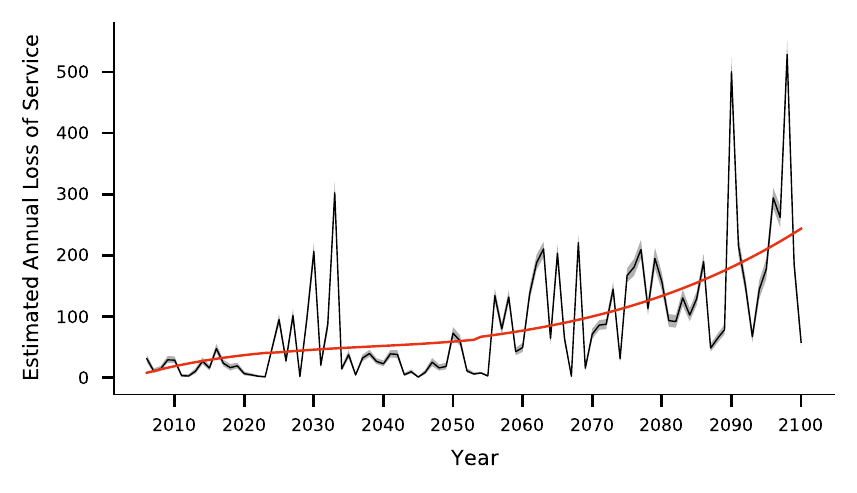}
    \caption{Time series of the estimated annual total LOS for the 2006-2100 period. The solid black line shows the expectations of the aggregated distributions. The shaded area shows the 5\% to 95\% range across the aggregated distribution. The solid red line is obtained by smoothing the solid black line with a Savitzky–Golay filter \cite{savgol_filter}. }
    \label{fig:annual_los}
\end{figure}

In addition, a time series of annual total LOS is constructed using the simulation results of the example days and the clustering analyses. This is done by duplicating the simulation results for each example day and assigning them to all days in the same cluster. The days are then grouped by year and the convolution of all non-zero distribution is computed to give the estimation of the annual total LOS. From Figure \ref{fig:annual_los}, the constructed time series data shows an overall upward trend regarding the estimated annual total LOS. Moreover, the spikes suggest that, in the climate model output used, there exist some extreme hot years with significantly more and hotter days. The overall upward trend suggests that the railway system's resilience to high-temperature-induced disruptions could be compromised under future climate scenarios.

\subsection{Shock-disruption relationship }

    \begin{figure*}[h]
         \centering
         
         \begin{subfigure}[b]{0.65\columnwidth}
             \centering
             \includegraphics[width=\columnwidth]{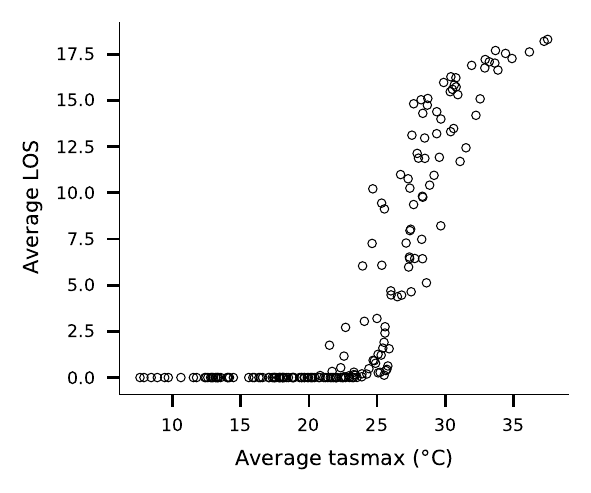}
             \caption{}
             \label{fig:G2a}
         \end{subfigure}
         \hfill
         \begin{subfigure}[b]{0.65\columnwidth}
             \centering
             \includegraphics[width=\columnwidth]{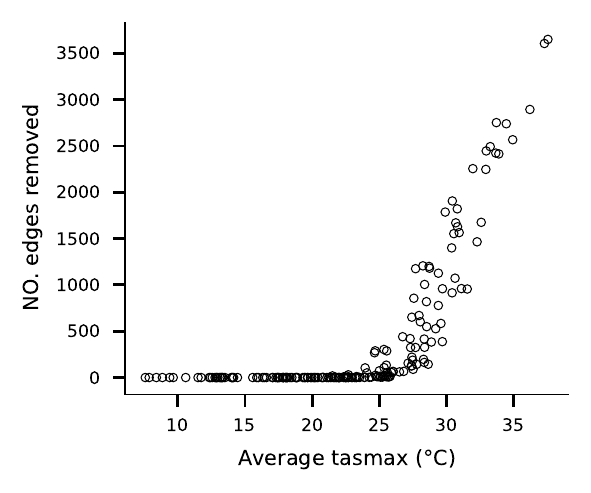}
             \caption{}
             \label{fig:G2b}
         \end{subfigure}
         \hfill
         \begin{subfigure}[b]{0.65\columnwidth}
             \centering
             \includegraphics[width=\columnwidth]{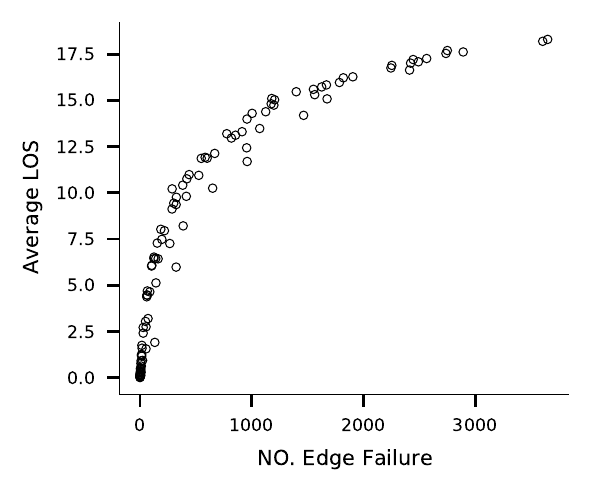}
             \caption{}
             \label{fig:G2c}
         \end{subfigure}
         
            \caption{Relationship between: (a) The average of measured cumulative loss of service, $\mathcal{LOS}$ and the average tasmax; (b) the expectation of the number of edge failures $\Psi$ and the average tasmax; (c) The average of measured cumulative loss of service, $\mathcal{LOS}$ and expectation of the number of edge failures, $\Psi$. The average tasmax is the average of the tasmax relevant to the asset layer.}
            \label{fig:G2}
    \end{figure*}
In this case study, the external weather condition is regarded as a form of external shock imposed on the railway system. The direct effect of such shock is the failure and removal of edges from the asset layer. Service disruption occurs when the railway system fails to deliver part of its service. Infrastructure systems are believed to be complex systems that often exhibit non-linearity. The railway system in this case study is no exception, as a non-linear relationship is observed between the intensity of external shock received and the severity of service disruption caused.

Figure \ref{fig:G2a} shows this shock-disruption relation, with the intensity of the shock measured by the national average of tasmax and the severity of the disruption indicated by the simulated LOS. The plot uses the simulation results of the 105 example days, whose expected number of failure, $\Psi$, is greater than 1. For each day, the national average of tasmax is plotted against the simulated cumulative loss of service. As the temperature increases, the increase of the consequential loss of services presents a threshold behaviour with an apparent surge as the temperature increase above 25 \textdegree C. Following the surge, there is a steady increase with the increasing temperature, which then shows a trend to plateau. This suggests the possibility that changes in the climate system can lead to some disproportional disruptions to the infrastructure systems.

This shock-disruption relationship is then further broken into a shock-damage-disruption relation. The shock-disruption relationship in Figure \ref{fig:G2a} is in fact a combined effects of the shock-damage relationship in Figure \ref{fig:G2b} and the damage-disruption relationship in Figure \ref{fig:G2c}. Figure \ref{fig:G2b} shows the relationship between the intensity of the shock and the extent of damage caused, where a surge happens at a 25 \textdegree C national average of tasmax followed by a steady increase with increasing temperature. Figure \ref{fig:G2c} shows the relationship between the extent of damaged caused and the severity of service disruption. The plot has a sharp increase as few edges are removed and then keeps on growing at a slower rate.

\subsection{Climate-based, random and targeted failures }
As mentioned in Section \ref{sec:1}, neither the random failure scenarios, where failures happen randomly across the network, nor the targeted attack strategy, where network component failures are targeted to simulate the most extensive possible disruption, may be able to capture the feature of weather-related disruptions. Therefore, two separate sets of simulations, one implementing the random failure scenarios and the other implementing the targeted attack strategy, are carried out to compare the effects of the these against climate-based disruption. For a given day, the expected number of edge failures, $\Psi$, is calculated using Equation \ref{eq:Psi}. Then, for the random strategy, in each Monte Carlo run, $\Psi$ number of edges are randomly selected and removed from the network, followed by the same rerouting and repairing efforts with LOS calculated using the same approach. For the targeted strategy, the first $\Psi$ edges with the most traffic are removed from the network, followed by the same assessment procedure. 

\begin{figure}[h]
        \begin{center} 
        	\includegraphics[width=1\columnwidth]{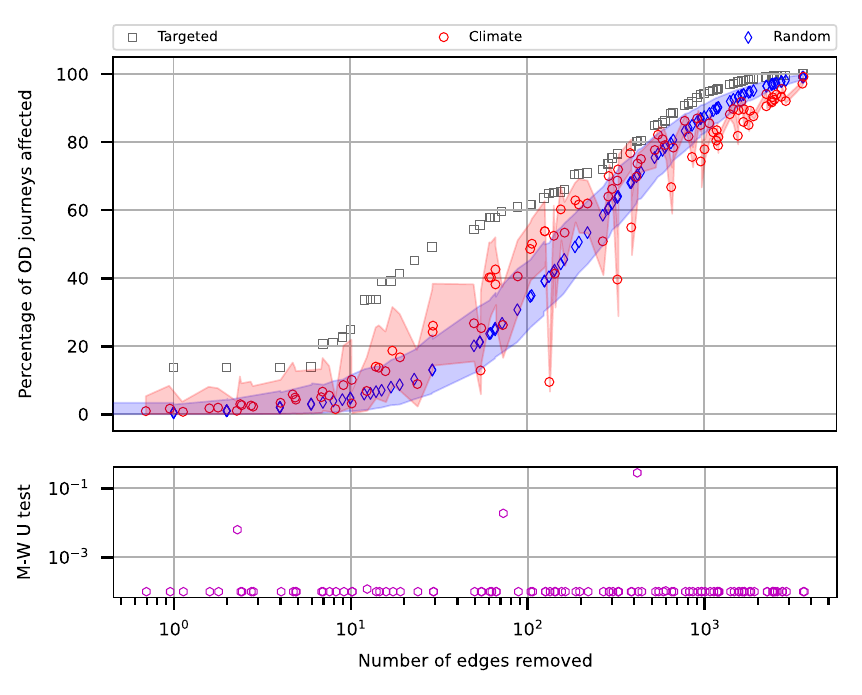}
        \end{center} 
        \caption{ upper): Percentage of OD journey interrupted at the onset of disruption without any rerouting or repairing effort against the number of the nodes removed. The sample size is 1000. The shaded area shows the 2.5\% to 97.5\% range. Markers are the means of each sample. lower): P-value of the Mann-Whitney U test between samples from climatic and random failure scenarios vs. the number of nodes removed. Any p-value smaller than 0.0001 is replaced with 0.0001.  } 
        \label{attack_eff}
    \end{figure}

Firstly, the percentage of OD flows interrupted at the onset of component failure without any rerouting or repair attempts is compared between the three strategies (Figure \ref{attack_eff}). When the same number of edges are removed from the network, the strategy that results in a higher percentage of OD disruption is believed to be more disruptive. The results show that climate-based strategy tends to sit between random and targeted. In particular, for the extent of the failed portion of the network edges intermediate between a sparse and a total number of failures, the climate-based strategy is more disruptive than random and less than targeted. For an extremely high number of edges involved, the disruption of a random strategy would exceed the one generated by the climate-based though there is only a slight difference between the three strategies. A Mann-Whitney \textit{U} test \cite{Mann_Whitney} is carried out between samples from the climatic-based and random failure scenarios. The results of this statistical test highlight how different climatic disruptions are from random disruptions. 

\begin{figure}[h]
        \begin{center} 
        	\includegraphics[width=1\columnwidth]{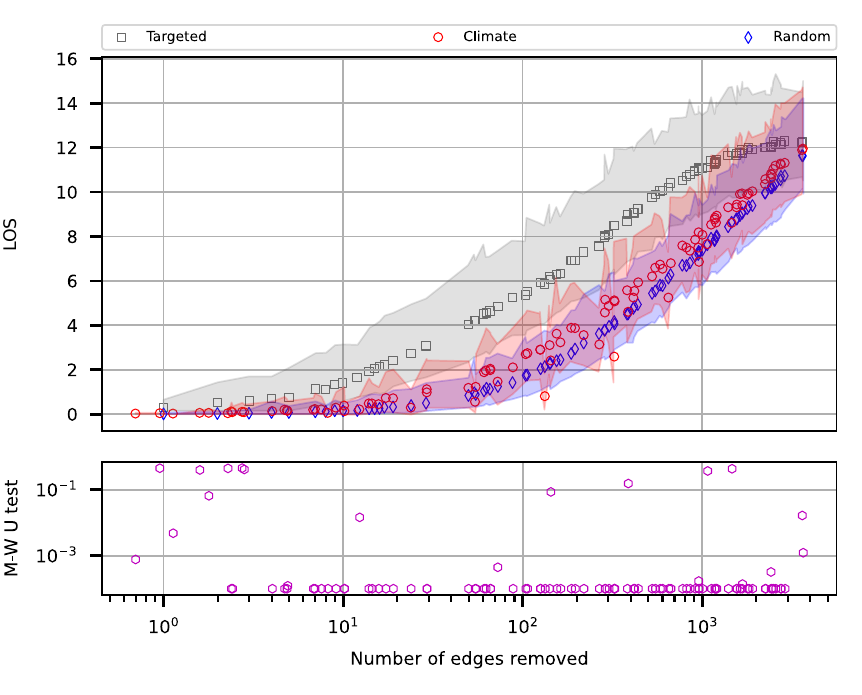}
        \end{center} 
        \caption{ upper)The measured LOS against the number of edges removed. The sample size is 250. The shaded area shows the 2.5\% to 97.5\% range. Markers are the means of each sample. lower) P-value of the Mann-Whitney U test between samples from climate-based and random failure scenarios against the number of nodes removed. Any p-value smaller than 0.0001 is replaced with 0.0001 for visualisation purposes.} 
        \label{res_Tri_area}
    \end{figure}

Figure \ref{res_Tri_area} shows the distribution of the measured cumulative loss of service against the number of edges removed for the three strategies. The results show that the climate-based failure scenarios cause more cumulative service loss than random in most cases. The Mann-Whitney U test shows a few cases where LOS sampled from the random and the climate-based strategies return no statistical difference. In contrast, the targeted strategy sits distinctly above both the random and the climate-based for all extents of edge failure. Overall, the system suffers a higher level of loss of service under the climate-based failure scenarios than the random one when the same number of edges are removed from the network. It means the system has a higher level of resilience toward random failure scenarios than climate-based. 

% \begin{figure}[p]
% \centering
% \includegraphics{fig1}
% \caption{Caption 1}
% \includegraphics{fig2}
% \caption{Caption 2}
% \end{figure}

\section{Discussion} \label{sec:4}

The case study points to three key findings: 1) There is an overall increasing trend in the estimated annual total LOS over the 95-year-period, where increased frequency and intensity of extreme events drive the average LOS upwards. 2) The severity of the disruptions caused, measured with the cumulative LOS, increases non-linearly with the increasing intensity of the external shock applied. 3) Random failure models tend to overestimate the network's resilience compared to climate-based failure scenarios. When climate-based disruptions are considered, the network function degradation is more severe than in the random failure scenarios. Nevertheless, it is statistically different from a network disrupted by malicious attacks. 

The direct use of climate model outputs also provide the possibility to assess the resilience of the infrastructure systems to different climate change scenarios and provides a more comprehensive assessment of the threat of climate change. This is achieved through two aspects. First, the introduction of the Monte Carlo simulation in the assessment framework provides a density distribution of the possible extent of service loss in the future instead of a single numerical value. It provides a sense of how broad the outcomes would fall and the associated likelihood of those outcomes. Second, it can exploit the tremendous number of dataset climate change research provides for future climate projections. One climate model output dataset can only provide one possible view into the future, and only a sufficient collection of outputs from different climate model simulation runs can give a more reliable prediction with the degree of uncertainty and error range addressed. Likewise, a more informative and reliable conclusion regarding the impact of climate change would be one that is drawn through the evaluation of assessed resilience to adequate sets of climate model outputs from different global and regional models with different model ensembles and initiation. 

The proposed approach initiates disruptive events in infrastructure systems using the output data from climate change research. Such an approach offers advances compared to the weather generation methods used in literature \cite{panteli2015influence,panteli_power_2017,fu2018integrated}. Those methods mostly involve some form of manipulation of present-day weather statistics and presume that future weather patterns remain the same as the current day. The proposed method overcomes such presumptions and can generate behaviourally realistic and physically viable climate hazards under future climate change scenarios. However, the proposed approach is limited by the spatial and temporal resolution of the climate change data. For example, the highest spatial resolution offered by EURO-CORDEX is \(\sim\)12km and the majority of the outputs are with a daily temporal resolution. In contrast, those weather simulators can offer up to half-hourly weather profiles. Although the case study here only requires daily weather profiles, future application of the method may requires weather profile of higher temporal resolution. 

The proposed failure initiation approach is applicable to any infrastructure system with most types of climate hazards, provided the fragility functions become available. By introducing multiple inter-dependent asset layers subjected to multiple climate hazards, the proposed approach has the potential to simulate cascading failures in an interconnected infrastructure network and assess the resilience of the system. Future applications of the approach should see case studies on interdependent infrastructure systems with compound climate hazards.

The case study also highlights two main challenges in implementing the proposed approach. First, a huge amount of computational power is required to carry out the Monte Carlo simulations in the proposed probabilistic resilience assessment. In the case study, a clustering analysis is carried out to limit the number of days input into the assessment framework. Doing so brings the amount of calculation within computational power available and provides important insights into the 95-year trend. The second challenge is the lack of empirical data and information on industrial practice for the specification of fragility functions and railway system dynamics. For example, the industrial practice of rerouting, the prioritisation of repairing, and the estimation of spare capacity. A few assumptions are made in the case study to the most general form. Provided such information become available, a more rigorous quantification of infrastructure resilience to future climate can be addressed.

\section{Conclusion} \label{sec:6}

This work presented a method to generate network failure scenarios and system disruptions using climate change research data for the assessment of the infrastructure system's climate resilience. The case study attempts to quantify the resilience of Great Britain's railway passenger transport system to high-temperature-related track buckling under the RCP8.5 climate change scenario. Findings from the case study support the two arguments that motivate this framework's proposal: 1) Random failure models tend to overestimate the network's resilience; 2) The system quality of services degrades non-linearly with the magnitude of the disruption. Together, they prove the need for linking climate change resilience assessment to system-level functional loss as opposed to a mere count of the failed nodes or the identification of a threshold for network fragmentation.

{\appendix[The Rerouting Algorithm]
The modified minimum-cost maximum-flow algorithm based on the Edmonds-Karp algorithm \cite{edmonds1972theoretical} used for the rerouting in the case study is shown in Figure \ref{fig:rerouting_alg}. The code involves a large amount of path search as there are 2,282,270 OD pairs in the system model with an asset layer constituting 2484 nodes and  4524 edges. The search for alternative paths is time-consuming and memory-intensive. To meet the limitation in computational power available, the algorithm is set to eliminate the alternative path search for any OD pair with fewer than 15 passenger trips or geographical path length less than 30km. Those trips can be regarded as trips that are likely to be aborted due to their relatively small demand figures or met by using alternative transportation services due to their short geographical distance. These include OD pairs with potentially no passengers in the days considered, which, together, contribute to 99\% toward OD pair count. By eliminating those OD pairs from the path search, the computational time can be reduced by 96\%. In fact, the sum of those OD pairs contributes 71\% toward total passenger trips. The non-rerouted demand is assumed to resume travel as the service is re-established on the original path. Up to that time, it counts as unsatisfied demand. A small-scale trial computation was carried out to assess the suitability of this approach. The difference in the absolute amount of flow delivered between a full rerouting and the reduced rerouting strategy chosen is less than 5\% in the vast majority of the cases (Figure \ref{fig:hist_reroute}).

\begin{figure}
    \centering
    \includegraphics[width=1\columnwidth]{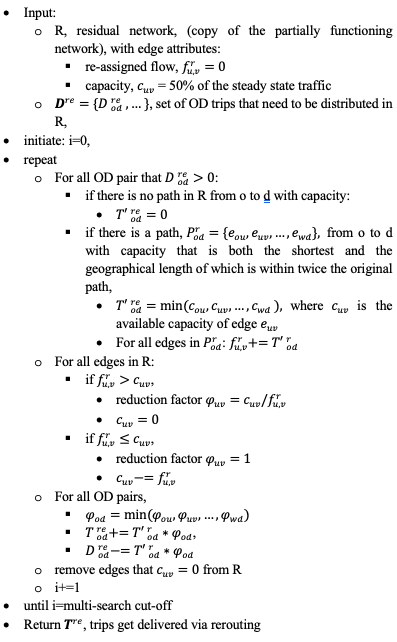}
    \caption{Psuedocode explaining the algorithm}
    \label{fig:rerouting_alg}
\end{figure}

\begin{figure}
    \centering
    \includegraphics[width=1\columnwidth]{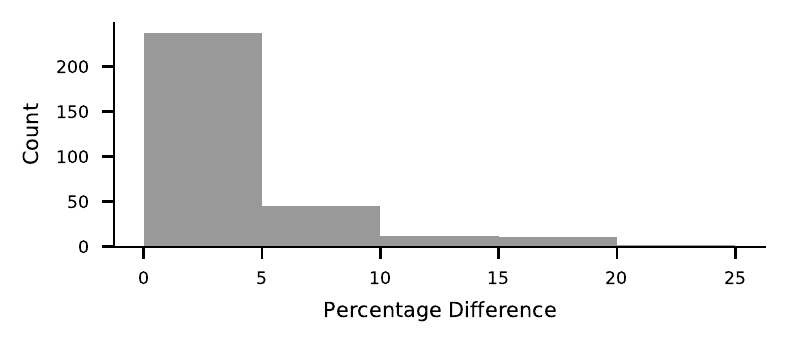}
    \caption{Histogram of the percentage difference between the amount of flow delivered at each time step between a full rerouting effort and the reduced rerouting effort. Twelve failure scenarios are included in the sample. The amount of trips delivered with the two rerouting efforts are compared at every time step.}
    \label{fig:hist_reroute}
\end{figure}
}

\bibliographystyle{IEEEtran}

\bibliography{QL/Q2_Art1.bib}

\end{document}